\documentclass[12pt,preprint]{aastex}
\usepackage{graphicx}
\usepackage{epstopdf}
\usepackage{natbib}
\newcommand{\oxy}{[\ion{O}{3}]~$\lambda$5007}
\newcommand{\nitro}{[\ion{N}{2}]~$\lambda$6584}
\newcommand{\halpha}{H$\alpha$}
\newcommand{\hbeta}{H$\beta$}
\newcommand{\hetwo}{\ion{He}{2}~$\lambda$4686}

\pdfoutput=1

\begin{document}

\shortauthors{Prieto et~al.}

\title{MUSE Reveals a Recent Merger in the Post-starburst Host Galaxy of
  the TDE ASASSN-14li}

\author{J.~L.~Prieto\altaffilmark{1,2},
  T.~Kr\"{u}hler\altaffilmark{3},  J.~P.~Anderson\altaffilmark{4},
  L.~Galbany\altaffilmark{2,5}, C.~S.~Kochanek\altaffilmark{6,7},
  E.~Aquino\altaffilmark{8}, J.~S.~Brown\altaffilmark{6},
  Subo~Dong\altaffilmark{9}, F.~F\"{o}rster\altaffilmark{10,2},
  T.~W.-S.~Holoien\altaffilmark{6,7,11}, H.~Kuncarayakti\altaffilmark{2,5},
  J.~C.~Maureira\altaffilmark{10}, F.~F.~Rosales-Ortega\altaffilmark{12},
  S.~F.~S\'anchez\altaffilmark{8}, B.~J.~Shappee\altaffilmark{13,14}, K.~Z.~Stanek\altaffilmark{6,7}}

\altaffiltext{1}{N\'ucleo de Astronom\'ia de la Facultad de Ingenier\'ia, Universidad Diego Portales, Av. Ej\'ercito 441, Santiago, Chile}
\altaffiltext{2}{Millennium Institute of Astrophysics, Santiago, Chile}
\altaffiltext{3}{Max-Planck-Institut f{\"u}r Extraterrestrische Physik (MPE), Giessenbachstrasse 1, 85748 Garching, Germany}
\altaffiltext{4}{European Southern Observatory, Alonso de C\'ordova
  3107, Casilla 19001, Santiago, Chile}
\altaffiltext{5}{Departamento de Astronom\'ia, Universidad de Chile, Camino El Observatorio 1515, Las Condes, Santiago, Chile}
\altaffiltext{6}{Department of Astronomy, Ohio State University, 140 West
18th Avenue, Columbus, OH 43210, USA}
\altaffiltext{7}{Center for Cosmology and AstroParticle Physics, The Ohio
State University, 191 W. Woodruff Ave., Columbus, OH 43210,
USA}
\altaffiltext{8}{Instituto de Astronom\'ia, Universidad Nacional
  Aut\'onoma de M\'exico, A.P. 70-264, 04510, D.F., M\'exico}
\altaffiltext{9}{Kavli Institute for Astronomy and Astrophysics, Peking
University, Yi He Yuan Road 5, Hai Dian District, Beijing
100871, China}
\altaffiltext{10}{Centro de Modelamiento Matem\'atico, Universidad de Chile, Av. Blanco Encalada 2120 Piso 7, Santiago, Chile}
\altaffiltext{11}{US Department of Energy Computational Science Graduate Fellow}
\altaffiltext{12}{Instituto Nacional de Astrof{\'i}sica, {\'O}ptica y Electr{\'o}nica, Luis E. Erro 1, 72840 Tonantzintla, Puebla, M\'exico}
\altaffiltext{13}{Carnegie  Observatories, 813 Santa Barbara Street,
  Pasadena, CA 91101, USA}
\altaffiltext{14}{Hubble and Carnegie-Princeton Fellow}

\begin{abstract}
We present MUSE integral field spectroscopic observations of the host
galaxy (PGC~043234) of one of the closest ($z=0.0206$, $D\simeq 90$~Mpc) and
best-studied tidal disruption events (TDE), ASASSN-14li. The MUSE
integral field data reveal asymmetric and filamentary structures that extend up 
to $\gtrsim 10$~kpc from the post-starburst 
host galaxy of ASASSN-14li. The structures are traced only through the 
strong nebular \oxy, \nitro, and \halpha\ emission lines. The total
off nuclear \oxy\ luminosity is $4.7\times
10^{39}$~erg~s$^{-1}$ and the ionized H mass is
$\rm \sim 10^4(500/n_e)\,M_{\odot}$. Based on the BPT diagram, the
nebular emission can be driven by either AGN photoionization or shock excitation, with AGN
photoionization favored given the narrow intrinsic line widths.  
The emission line ratios and spatial distribution strongly resemble
ionization nebulae around fading AGNs such as IC~2497 (Hanny's
Voorwerp) and ionization ``cones'' around Seyfert~2 nuclei. The
morphology of the emission line filaments strongly suggest that
PGC~043234 is a recent merger, which likely triggered a strong starburst 
and AGN activity leading to the post-starburst spectral signatures 
and the extended nebular emission line features we see today. We briefly discuss the
implications of these observations in the context of the strongly enhanced
TDE rates observed in post-starburst galaxies and their connection to
enhanced theoretical TDE rates produced by supermassive black-hole binaries. 
\end{abstract}


\section{Introduction}

When a star passes within the tidal radius of a supermassive
black-hole (SMBH), the strong gravitational tidal forces can tear it apart, potentially
producing a short-lived luminous flare known as a tidal disruption
event \citep[TDE;][]{rees88,evanskochanek89,strubbe09}. TDEs
can be used to find SMBHs in the centers of
galaxies, to study their mass function and its evolution in cosmic time,
to study the properties of the disrupted star and the stellar debris, and to
determine the SMBH mass and spin \citep[see review by][and references
therein]{komossa15}. The theoretical TDE rate in a galaxy with a single SMBH 
has been estimated to be $10^{-5} - 10^{-4}$ per year
\citep[e.g.,][]{wangmerritt04,stonemetzger16}. 
However, if a galaxy or galaxy merger has created a SMBH binary, the
TDE rate  could increase by a few orders of magnitude
\citep[e.g.,][]{chen09,chen11}. The TDE rate can also
be higher in steep central stellar cusps
\citep[e.g.,][]{magorriantremaine99,stonevanvelzen16}.

In the last few years, $\gtrsim 50$ TDE candidates\footnote{See
  {\tt https://tde.space}.} have been
identified at different wavelengths from $\gamma$-rays to optical
\citep[e.g.,][]{esquej07,gezari08,arcavi14,holoien14ae,holoien14li,holoien15oi}. One
of the most unexpected observational results from these discoveries
was the finding by \citet{arcavi14} that a large fraction of TDEs have post-starburst
hosts, galaxies with strong Balmer absorption lines (A-type stellar
populations with ages $\sim 100-1000$~Myr) on top of a spectrum
characteristic of an old (elliptical) stellar population with
weak or no evidence for very recent star-formation (E+A galaxies;
e.g., \citealt{zabludoff96}). In a follow-up study using SDSS galaxies and a
larger TDE host sample, \citet{french16} confirm this result and
estimate that the observed TDE rate might be more than two orders of
magnitude higher in galaxies with strong Balmer absorption features
compared to normal galaxies. Given this puzzling observation and the
theoretical expectation of enhanced TDE rates in galaxies with
SMBH binaries, detailed studies of TDE host galaxies could provide important
insights into the physical mechanism that produces this rate enhancement. 

The All-Sky Automated Survey for SuperNovae
\citep[ASAS-SN;][]{shappee14}, a transient survey of the whole sky at
optical wavelengths, has discovered three of the closest
($90-220$~Mpc) and best studied TDEs (ASASSN-14ae,
\citealt{holoien14ae,brown16a}; ASASSN-14li, \citealt{holoien14li};
ASASSN-15oi, \citealt{holoien15oi}), providing
an excellent sample for detailed host galaxy studies. In particular,
ASASSN-14li at $z=0.0206$ ($D\simeq 90$~Mpc), discovered in November 2014, 
has been the best studied TDE to date at all wavelengths, including
the optical/UV \citep{holoien14li,cenko16,brown16b}, radio
\citep{alexander16,vanvelzen16,romero16}, X-rays \citep{holoien14li,miller15},
and mid-infrared \citep{jiang16}, as well as with theoretical
modeling \citep{krolik16}. The archival, nuclear SDSS
spectrum of the host galaxy of ASASSN-14li, PGC~043234
(VIII Zw 211; $\rm M_{\star}\simeq 3\times 10^9$~M$_{\odot}$), shows strong 
Balmer lines in absorption and no strong evidence for current
star-formation \citep{holoien14li}. Indeed, the galaxy has the
highest Lick H$\delta_{\rm A}$ index in the TDE host sample studied by 
\citet{french16}, indicating a strong post-starburst stellar
population in its nuclear region with an age of $\sim 100$~Myr. 

In this letter, we present integral field spectroscopic observations
of the host galaxy  of ASASSN-14li and its surroundings obtained in
early 2016. In Section~\S\ref{sec2} we discuss the observations and data
reduction. In Section~\S\ref{sec3} we present the results and the analysis
of the data. We discuss our results in
Section~\S\ref{sec4}. Throughout the paper, we assume a distance to
ASASSN-14li of $D=90.3$~Mpc corresponding to a linear scale of
$0\farcs 44$/kpc \citep{holoien14li}.

\section{Observations and Data Reduction}
\label{sec2}

We observed the field of ASASSN-14li as part of the All-weather MUse
Supernova Integral field Nearby Galaxies (AMUSING; \citealt{galbany16}) 
with the Multi Unit Spectroscopic Explorer (MUSE, \citealt{bacon10}) on
ESO's Very Large Telescope UT4 (Yepun). MUSE is a state-of-the-art integral field
spectrograph with a field of view of $1\,\mathrm{arcmin}^2$ and
$0\farcs 2$ spaxels. It covers the spectral range $4800-9300$~\AA\ with a
resolving power of  $R \simeq 1800-3000$. Our MUSE data were obtained
on 2016-01-21 and consisted of four dithered exposures each with an
integration time of 698~sec. The sky conditions were clear, and we
measure a full-with-half-maximum for the stellar point-spread function 
of 1\farcs{05} at 5000\,\AA\, and 0\farcs{85} at 9000\,\AA, respectively. 

We reduced the MUSE spectroscopy with version 1.2.1 of the pipeline
provided by ESO \citep{weilbacher14}, which applies a
bias, flat-field, illumination, field geometry correction and
background subtraction. The wavelength solution was tied to arc lamp
frames, refined using skylines in the science data and converted into
the heliocentric reference frame. Observations of the spectrophotometric
standard GD108, taken immediately after the science data, provided the
flux calibration. The final product of this reduction process is a
data cube with two celestial coordinates sampled at 0\farcs{2} 
and one wavelength dimension sampled at 1.25\,\AA. 
We checked the astrometric zeropoint of the MUSE data using the SDSS DR12
\citep{alam15} $r$ and $i$-band images of the field. We found and
corrected astrometric shifts of 1\farcs{03} in RA and 1\farcs{33} in Dec
with respect to the SDSS images. 

\section{Results and Analysis}
\label{sec3}

\begin{figure}
\leftskip-2cm
\epsscale{1.2}
\plotone{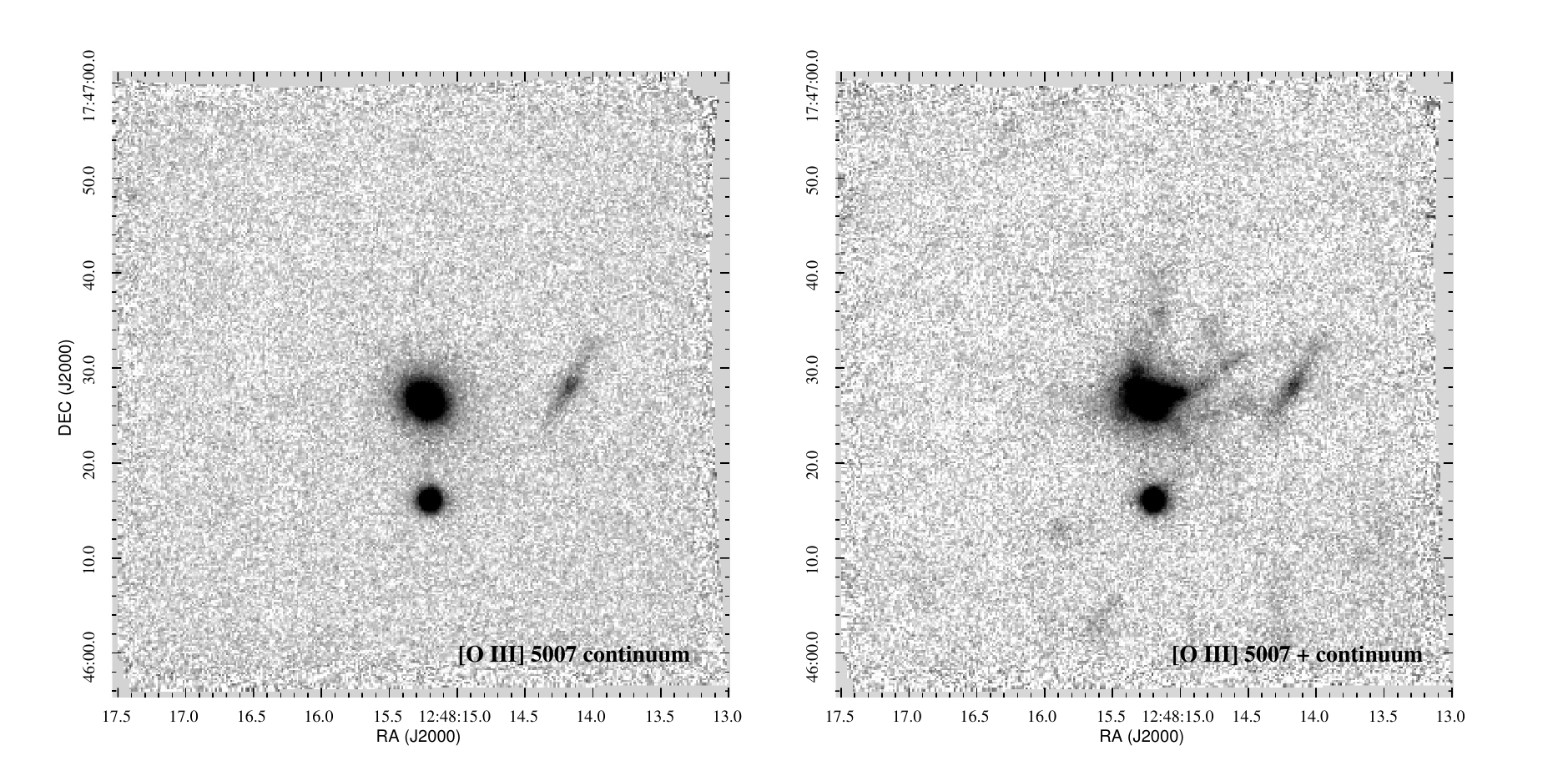}
\caption{Grey scale images showing the full $1'\times1'$ MUSE field 
  centered on PGC~043234 (RA=12:48:15.24, DEC=+17:46:26.5), the host galaxy of the nearby TDE
  ASASSN-14li. The {\it left} panel shows an image of the
  continuum emission at 5100~\AA, just to the blue of the strong \oxy\
  nebular line at the redshift of PGC~043234. The {\it right} panel
  shows an image at 5110~\AA\, which includes
  \oxy\ nebular line emission at the redshift of
  PGC~043234, that clearly reveals the presence of extended \oxy\
  emission. The extended source to the west of PGC~043234 
   is a background edge-on galaxy at $z=0.15$ (RA=12:48:14.18,
   DEC=+17:46:28.0) and the point source to the south of PGC~043234 is
   a foreground Galactic star (RA=12:48:15.21, DEC=+17:46:16.06).}
\label{fig1}
\end{figure}

An initial inspection of the MUSE data immediately showed the presence
of extended line emission around PGC~043234 that was not detected in
the SDSS $ugriz$ images. The left panel of 
Figure~\ref{fig1} shows a full $1\arcmin \times 1\arcmin$ image of the
continuum just to the blue of the nebular \oxy\ emission line (at
$\sim 5100$~\AA\ in the observer's frame) and the right panel shows an
image including the nebular \oxy\ emission line and its underlying
continuum (at $\sim 5110$~\AA\ in the observer's frame). The extended
source to the West of PGC~043234 is an edge-on background galaxy at
$z=0.1517$, measured using nebular emission lines from the galaxy, and
the bright point-source to the South is a Galactic star. The extended 
features around PGC~043234 have strong \oxy\ in emission, and 
further analysis also reveals strong \nitro\ and \halpha\ emission at 
the same locations. 

Figure~\ref{fig2} shows a false color image of the MUSE field around
ASASSN-14li/PGC~043234. In blue is the continuum-subtracted 
\oxy\ emission line image, in green is the continuum-subtracted 
\nitro\ emission line image, and in red is the $i$-band continuum
flux image obtained from synthetic photometry of the MUSE data cube.  
We observe complex filamentary and asymmetric structures in the nebular emission
around PGC~043234 that resemble a galaxy merger that is undetected in the
continuum image. The emission extends to a projected distance of
$\gtrsim 5$~kpc ($\gtrsim 11 \arcsec$) from the nucleus of PGC~043234,
with some isolated nebular emission regions at $\gtrsim 10$~kpc
($\gtrsim 23 \arcsec$) from the nucleus. The ``arm'' structure to the
North-West of PGC~043234 extends to $\sim 5$~kpc, with a resolved extended peak in
nebular emission at $\sim 1.2$~kpc. There is also an extended region
of strong nebular emission $\sim 2$~kpc to the East/North-East of  
PGC~043234 with multiple emission peaks. 

The circles in Figure~\ref{fig2} define $1\arcsec$ radius apertures for
regions with strong nebular emission that we will study in more detail. 
In order to extract the fluxes of the nebular emission lines free from
the contamination of the stellar continuum, we modeled the
stellar continuum in the spectrum of each region using stellar population synthesis
models from STARLIGHT \citep{cidfernandes09} and following the
prescriptions presented in \citet{galbany14}. The \hbeta\ emission line is
undetected in most of the regions and we estimate 3$\sigma$ upper
limits on the emission line fluxes following \citet{shappee13}. 

In Table~\ref{tab1} we present the measured properties of the regions
defined in Figure~\ref{fig2}, including the coordinates, projected distances (in
kpc) and radial velocity differences (in km~s$^{-1}$, measured from
the \nitro\ line) with respect to the nucleus of PGC~043234, along
with integrated line luminosities (corrected for Galactic extinction) for
the strongest nebular emission lines (\oxy, \hbeta, \halpha, and \nitro).  
The nebular emission line luminosities of the nuclear region of PGC~043432 shown in
Table~\ref{tab1} are also included after subtracting the best-fit STARLIGHT model
for the absorption lines. The \halpha\ and \hbeta\
emission line luminosities for the nuclear region should be interpreted
with caution given the very strong stellar absorption and the
contribution from ASASSN-14li to the nuclear fluxes even 
at late epochs after its discovery \citep{brown16b}. 

In Figure~\ref{fig3} we show the integrated spectra of the regions
defined in Figure~\ref{fig2} with the highest \oxy\ line
luminosities. In addition to  H$\alpha$, H$\beta$,
[\ion{O}{3}], and [\ion{N}{2}], the \hetwo\ and [\ion{S}{2}]~$\lambda\lambda$6716,6731 nebular emission lines are
clearly detected in the region NW1. From the detection of the [\ion{S}{2}]
doublet we can estimate the electron density of the nebula
\citep{osterbrock06}. Using the {\tt temden} task in IRAF, the measured
[\ion{S}{2}]~$\lambda\lambda$6716,6731 flux ratio of 1.04, and an
assumed electron temperature of $\rm T_e=10^4$~K, the 
estimated electron density for the region is $\rm n_e \sim
500$~cm$^{-3}$. This implies a recombination timescale of $\rm
t_{rec}\sim 200$~yr \citep{osterbrock06}. Note that the
[\ion{S}{2}]~$\lambda$6731 emission line lies at the wavelength of the
telluric O$_{2}$ B-band absorption, making this estimate of the
density prone to significant systematic uncertainties. 

In Figure~\ref{fig4} we show the standard Baldwin-Phillips-Terlevich
\citep[BPT;][]{BPT} emission line diagnostics
diagram along with the [\ion{N}{2}]/H$\alpha$ and [\ion{O}{3}]/H$\beta$
flux ratios (in logarithmic scale) for all the nebular emission
regions defined in Figure~\ref{fig2} and the nucleus of
PGC~043234. The dashed and dotted lines in Figure~\ref{fig4} separate the
main sources of photoionization of nebulae between star formation (starburst) and AGN 
activity as defined in \citet[dashed line]{kewley01} and
\citet[dotted line]{kauffmann03}. Shock excitation can also produce
line ratios in the range of the BPT diagram between star-formation and AGN
\citep[e.g.,][]{allen08,rich11,alatalo16}. The area inside the magenta
region defined in Figure~\ref{fig4} contains values of the line ratios
that can be explained by shock models \citep{alatalo16}. 

All the \oxy\ and \nitro\ emitting line regions around PGC~043234,
including its nucleus, are consistent with photoionization by an AGN,
instead of current star-formation, and most of the values of the ratios are also
consistent with shock models. Given the presence of late-time Balmer
line contamination from TDE emission in the nucleus \citep{brown16b}, we
also measured the line ratios using the archival SDSS spectrum of the
host galaxy (obtained in 2007) using the same approach to subtracting the
strong stellar absorption lines. The resulting SDSS emission line fluxes
of the nuclear region are \mbox{log([\ion{N}{2}]/H$\alpha$)\,$\simeq$\,0.19} and
\mbox{log([\ion{O}{3}]/H$\beta$)\,$\simeq$\,0.38},  which still puts
the nucleus in the AGN/shock region of the BPT diagram (see
Figure~\ref{fig4}).  

\section{Discussion}
\label{sec4}

\begin{figure}
\vspace*{-5cm}
\epsscale{1.1}
\plotone{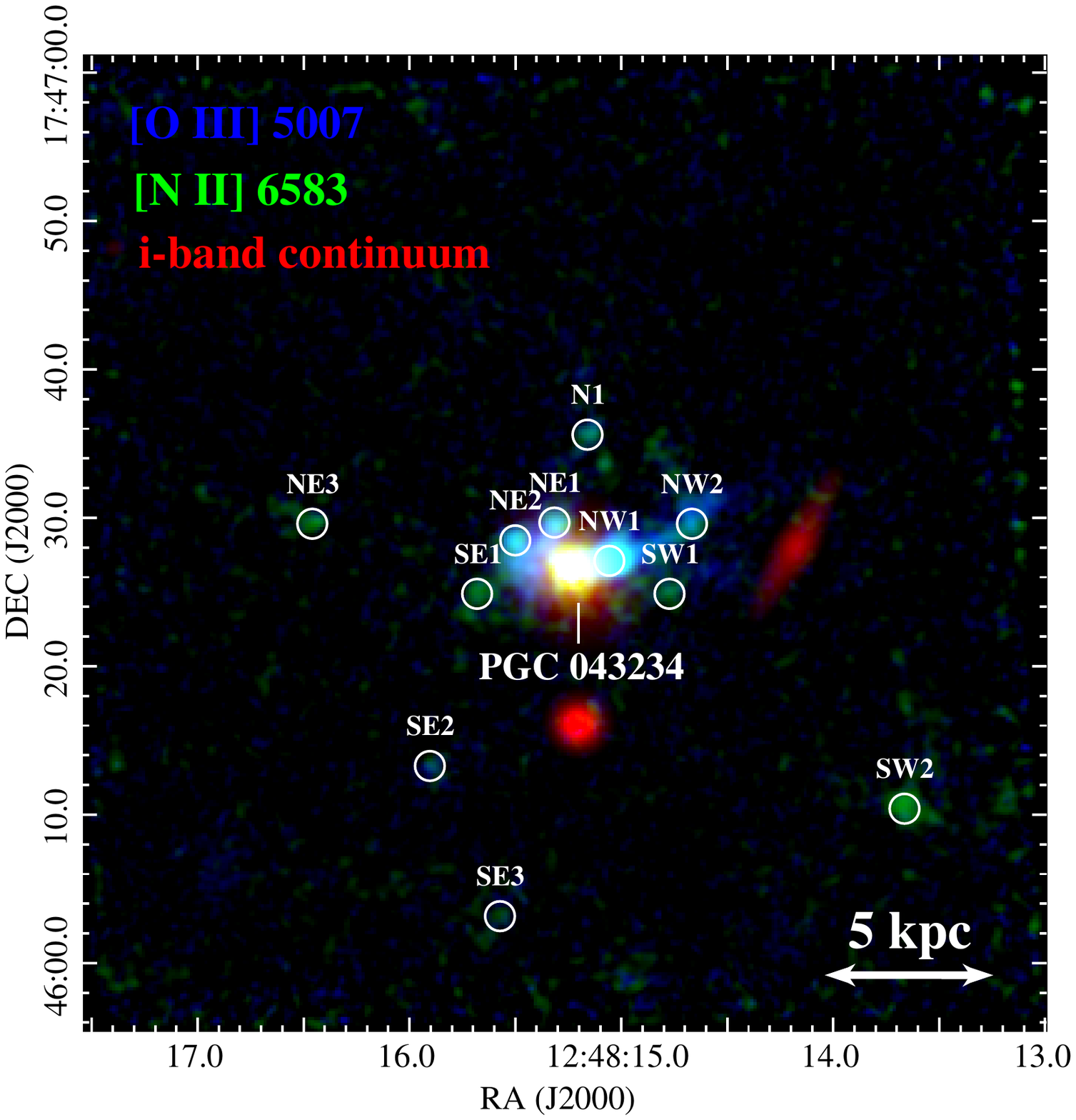}
\vspace*{-4cm}
\caption{False color image showing the integrated, continuum-subtracted nebular emission of 
 the \oxy\ (blue) and \nitro\ (green) lines at the redshift of PGC~043234
  (center). The continuum emission (red) is traced by the SDSS $i$-band flux obtained from
  synthetic photometry. The \oxy\ and \nitro\ nebular
  line emission is distributed in the nucleus of PGC~043234 and around it,
  with large filamentary structures extending to $\sim 12$~kpc from the
  nucleus. The strongest \oxy\ emission is $\sim 1.2$~kpc to the
  North-West (region NW1) of the nucleus. The 1$\arcsec$ radius
  circles with names are nebular emission regions defined for further
  analysis (see Table~\ref{tab1}).}
\label{fig2}
\end{figure}

\begin{figure}
\epsscale{1.0}
\plotone{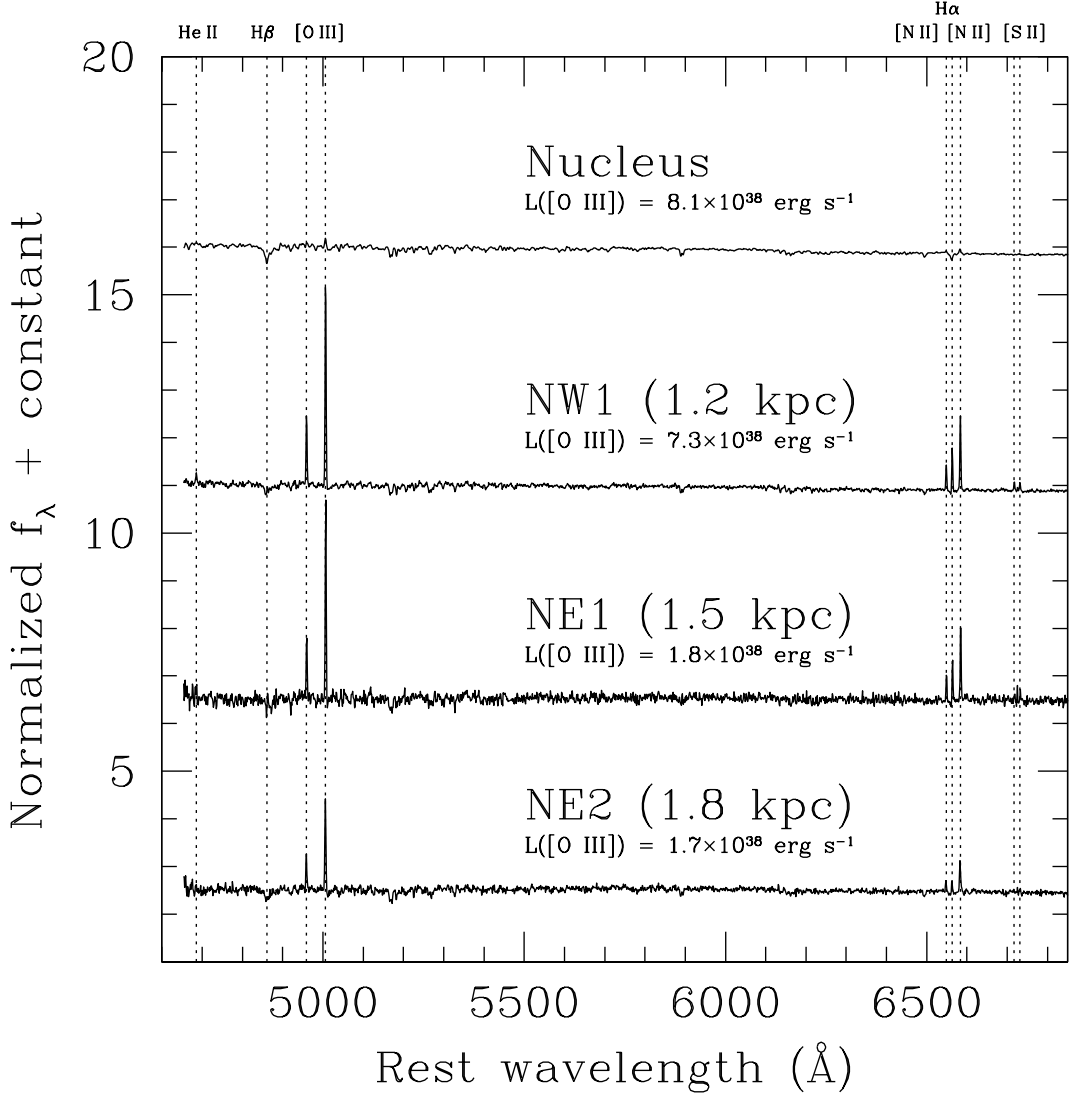}
\vspace*{-4cm}
\caption{Integrated spectra ($1\arcsec$ aperture radius) of the strongest \oxy\ emission
  regions identified in Figure~\ref{fig2} and the nucleus of 
  PGC~043234. Projected distances of each region from the nucleus are given in
  parenthesis. The main nebular emission lines in the spectra are
  [\ion{O}{3}]~$\lambda\lambda$4959,5007,  [\ion{N}{2}]~$\lambda\lambda$6548,6584, and
 \halpha. Weaker nebular lines detected in NW1 are \hetwo,
  \hbeta\ (also in absorption), and [\ion{S}{2}]~$\lambda\lambda$6716,6731. The dotted vertical lines show
  the rest wavelengths of the nebular emission lines. The spectra have
  been normalized by the local continuum at the \oxy\ line
  and shifted in flux for clarity. \label{fig3}
}
\end{figure}

We have presented MUSE integral field spectroscopic observations of
the nearby TDE ASASSN-14li host galaxy, PGC~043234, and its
environment ($\rm 26\,kpc \times 26\,kpc$). These data reveal the
presence of asymmetric and filamentary emission line 
structures that extend many kpc from the post-starburst host galaxy of 
ASASSN-14li (Figure~\ref{fig2}). The extended ($\gtrsim 5$~kpc)
filamentary structures are traced only by the detection of strong nebular
emission lines of \oxy,  \nitro, and \halpha, and are undetected in the
continuum, as illustrated in Figure~\ref{fig2}. The total off nuclear
line luminosities are $4.7\times 10^{39}$~erg~s$^{- 1}$, $1.8\times 10^{39}$~erg~s$^{-1}$, 
and $1.3\times 10^{39}$~erg~s$^{-1}$ for [\ion{O}{3}], [\ion{N}{2}],
and H$\alpha$, respectively, which implies an off-nuclear ionized H mass of $\rm M_{ion}
\sim 10^4(500/n_e)\,M_{\odot}$ \citep{osterbrock06}. 

\begin{figure}
\epsscale{1.0}
\plotone{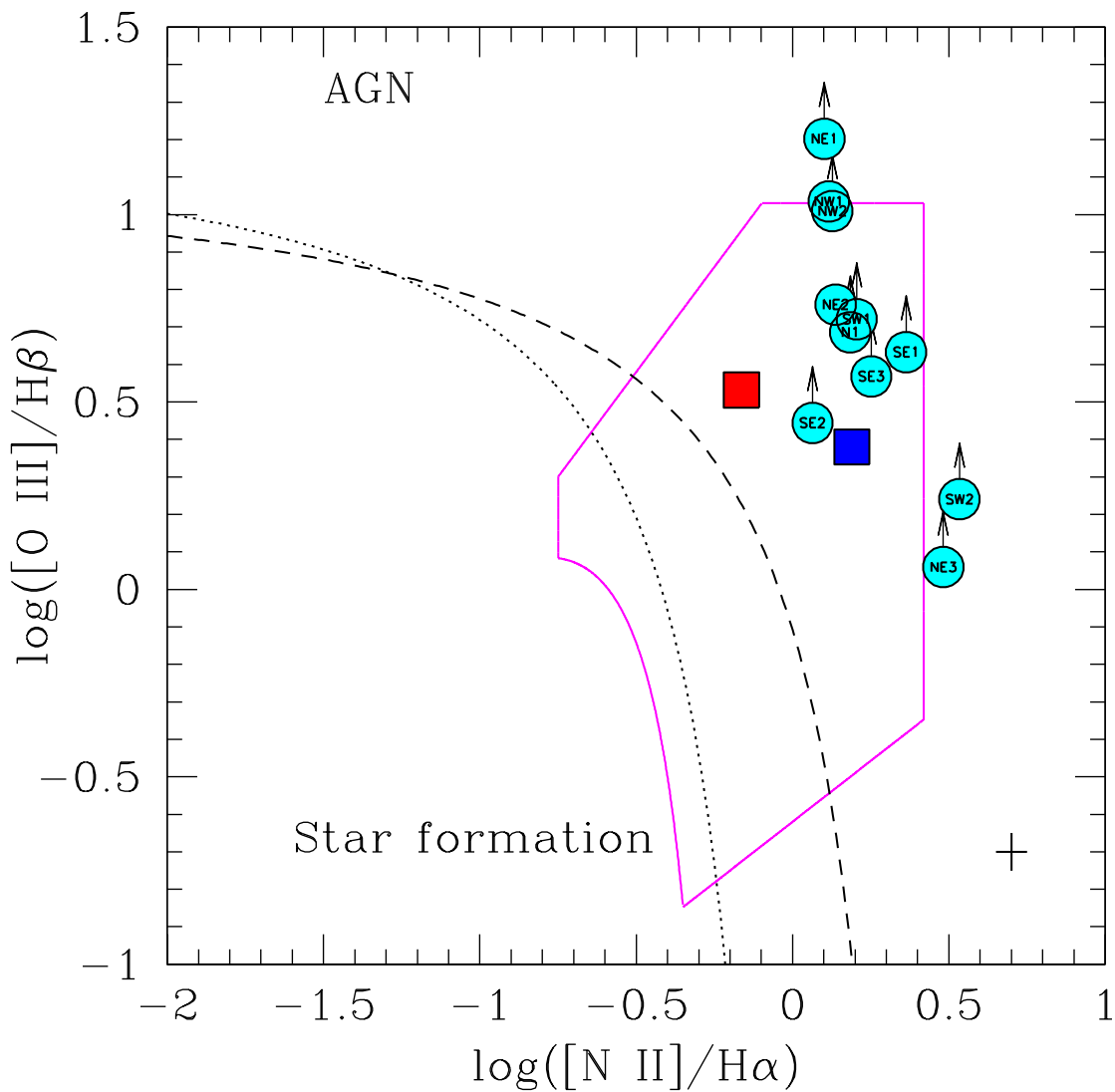}
\vspace*{-4cm}
\caption{Nebular emission line ratio diagnostic (BPT; \citealt{BPT}) diagram for the
  nucleus of PGC~043234 (red filled square from MUSE and blue filled
  square from the SDSS 2007 archival spectrum) and the off-nuclear
  nebular line emission regions identified in Figure~\ref{fig2} (cyan filled circles). The
  dashed and dotted lines separate the dominant photoionization
  mechanisms between star-formation and AGN, as shown in \citet{kewley01}
  and \citet{kauffmann03}, respectively. The line ratios inside the magenta
  region can also be explained by shocks \citep{alatalo16}. Average
  1$\sigma$ errors in the line ratios are shown in the lower right.} \label{fig4}
\end{figure}

The location of the main nebular emission line ratios for both the
extended emission line regions and the nucleus of PGC~043234 in the
BPT diagnostic diagram (Figure~\ref{fig4}) are consistent with
photoionization by an AGN or shock excitation, but not photoionization by current
star-formation. We do not favor shock excitation models as an explanation for 
the line ratios because the nebular emission lines have a low 
average intrinsic velocity dispersion of $\sim 40$~km~s$^{-1}$
and no broad wings. Fast shocks ($v \gtrsim 200$~km~s$^{-1}$) that can produce
the observed line ratios seem incompatible with the
line widths \citep{allen08}, while slow shocks ($v \lesssim
200$~km~s$^{-1}$) which might be compatible, but still broader, with
the line widths do not reproduce the line ratios \citep{rich11}. Also,
the \hetwo\ to \hbeta\ line ratio of 0.6 in the NW1 region is higher 
than in most of the fast shock models \citep{allen08}. 
We therefore conclude that the emission lines are most likely
photoionized by an AGN.

There are two lines of evidence that PGC~043234 was a weak AGN
prior to the TDE. The first, discussed in Section~\S\ref{sec3},
is that the stellar population-corrected SDSS archival spectrum from 
2007 of the nuclear region has line ratios suggesting AGN
activity. PGC~043234 is also associated with an unresolved FIRST \citep{becker95}
radio source with a luminosity of  $\rm L_{1.4\,GHz} \simeq
2.6\times 10^{21}$~W~Hz$^{-1}$ \citep{holoien14li} or $\rm \nu \times
L_{\nu} \sim 4\times 10^{37}$~erg~s$^{-1}$ that is typical of
low-luminosity AGN \citep[e.g.,][]{ho99}. Although the radio
luminosity could be produced by star formation, there is no evidence
for current star formation in the host galaxy from either the
SDSS/MUSE spectra or the overall spectral energy distribution
\citep{holoien14li}. However, the upper limit on the soft X-ray luminosity of PGC~043234
from the ROSAT All-Sky Survey \citep{voges99} of $\rm L_X < 6 \times
10^{40}$~erg~s$^{-1}$ \citep{holoien14li,miller15}, implies
an ionizing luminosity which is too small to explain the
extended emission line features. Using the \halpha\ luminosity of the
brightest off-nuclear emission region (NW1), assuming case B
recombination and following \citet{keel12}, we estimate a minimum required ionizing luminosity 
from a central source of $\rm L_{ion}\gtrsim 2\times 10^{41}$~erg~s$^{-1}$.

The production of strong emission lines can then be explained in two ways.
First, the pre-TDE nucleus could be a Seyfert~2, with other lines
of sight being exposed to much higher ionizing fluxes. Such
ionization ``cones'' are observed around local Seyfert~2
AGN on similar physical scales \citep[e.g.,][]{wilson94,keel12}.
Second, the observed emission line structures also resemble Hanny's Voorwerp, a large 
ionization nebula located $15-25$~kpc from the galaxy IC~2497 \citep{lintott09},
and other ionization nebulae where the line emission is thought to be
an echo of AGN activity in the recent past rather than a reflection of 
present day activity \citep[e.g.,][]{keel12,schweizer13}.

The [\ion{S}{2}] doublet line ratio measured in the spectrum of the
strongest \oxy\ emitting region (NW1) implies a recombination time
that is short compared to the light travel time to the furthest
emission line regions ($\sim 10^2$~years versus $\sim 10^4$~years),
which suggests that the pre-TDE source would more likely be a
Seyfert~2 rather than a Voorwerp. The short recombination timescale is
also at odds with a ``fossil nebula'' interpretation
\citep{binette87}. However, the possible systematics
errors associated with the density estimate from the [\ion{S}{2}]
doublet line ratio, due to the dominant telluric absorption correction
at that wavelength, makes this conclusion fairly uncertain. 
Also, the distribution of the emission line regions around
PGC~043234 do not clearly favor the geometry seen in 
Seyfert~2 ionization cones \citep[e.g.,][]{wilson94}.  

In either case, the overall morphology of the emission line features strongly
indicates that PGC~043234 recently underwent a merger, leaving
relatively dense gas on large scales with no associated stars. This is
consistent with both recent AGN activity and the post-starburst
spectrum of the galaxy, strongly supporting the galaxy-galaxy merger scenario
proposed for E+A galaxies \citep{zabludoff96,goto05}. The stellar
continuum emission itself is quite smooth, suggesting that we are observing
the merger at a relatively late time \citep[e.g.,][]{hopkins08}. In these late phases,
we might expect a relatively compact SMBH binary 
in the nucleus of the host galaxy, which would then
naturally produce a greatly enhanced TDE rate \citep[e.g.,][]{chen09,chen11}.
This would be an exciting possibility for explaining the 
high TDE rates that appear to be associated with post-starburst
galaxies \citep{arcavi14,french16}, including the host of
ASASSN-14li. Indeed, the detection of a radio source at $\sim 2$~pc
from the nucleus of PGC~043234 in high resolution EVN observations 
could be explained by a companion AGN \citep{romero16}.   

\acknowledgments

We thank Rick~Pogge for valuable discussions. Support for
JLP, LG, FF, HK  are provided by FONDECYT through the
grants 1151445, 3140566, 11130228, and 3140563, respectively, and by
the Ministry of Economy, Development, and Tourism's Millennium Science Initiative
through grant IC120009, awarded to The Millennium Institute of
Astrophysics, MAS. TK is supported through the S. Kovalevskaja Award
to P. Schady. CSK and KZS are supported by NSF grants AST-1515876
and AST-1515927. SD is supported by Grant No. XDB09000000
from the Chinese Academy of Sciences and Project 11573003 supported by
NSFC. TW-SH is supported by the DOE CSGF grant
DE-FG02-97ER25308. BJS is supported by NASA through Hubble
Fellowship grant HST-HF-51348.001. FF and JCM acknowledge support
from Basal Project PFB–03 and from Conicyt through the infrastructure
Quimal project No. 140003. Based on observations made with ESO
telescopes at Paranal Observatory under programme 096.D-0296(A).

\newpage


\begin{deluxetable}{lcccccccc}
\tabletypesize{\scriptsize}
\tablecolumns{9}
\tablewidth{0pt}
\tablecaption{Properties of the emission line regions defined in Figure~\ref{fig2}.}
\tablehead{
\colhead{Location} &
\colhead{RA} &
\colhead{DEC} & 
\colhead{$\rm \Delta r$} &  
\colhead{$\rm \Delta v$} & 
\colhead{L(\hbeta)} & 
\colhead{L([O~III]~$\lambda$5007)} &
\colhead{L(\halpha)} &
\colhead{L([N~II]~$\lambda$6584)} \\ 
\colhead{} &
\colhead{(J2000.0)} &
\colhead{(J2000.0)} & 
\colhead{(kpc)} &  
\colhead{(km~s$^{-1}$)} & 
\colhead{($10^{37}$~erg~s$^{-1}$)} & 
\colhead{($10^{37}$~erg~s$^{-1}$)} &
\colhead{($10^{37}$~erg~s$^{-1}$)} &
\colhead{($10^{37}$~erg~s$^{-1}$)} 
}
\startdata
Nucleus & 12:48:15.24 & +17:46:26.5 & 0.0 & 0.0 & 23.7 & 80.7  & 57.3 & 39.4 \\ 
NW1 & 12:48:15.06 & +17:46:27.1 & 1.2 & +48 & 6.8 & 73.4  & 18.4 & 24.0 \\ 
NE1 & 12:48:15.32 & +17:46:29.7 & 1.5 & +12 & $< 1.1$ & 17.7  & 4.9 & 6.1 \\ 
NE2 & 12:48:15.50 & +17:46:28.5 & 1.8 & +88 & 3.0 & 17.2  & 4.6 & 6.4 \\ 
SE1 & 12:48:15.68 & +17:46:24.9 & 2.8 & $-20$ & $< 0.9$ & 3.8  & 1.1 & 2.6 \\ 
SW1 & 12:48:14.77 & +17:46:24.9 & 3.0 & $-35$ & $< 0.8$ & 4.2  & 1.4 & 2.2 \\ 
NW2 & 12:48:14.67 & +17:46:29.6 & 3.8 & +43 & $< 1.0$ & 10.2  & 2.9 & 3.9 \\ 
N1 & 12:48:15.16 & +17:46:35.6 & 4.0 & $-9$ & $< 1.1$ & 5.3  & 1.8 & 2.7 \\ 
SE2 & 12:48:15.90 & +17:46:13.3 & 7.1 & $-20$ & $< 0.9$ & 2.5  & 0.9 & 1.1 \\ 
NE3 & 12:48:16.46 & +17:46:29.6 & 7.8 & $-4$ & $< 1.2$ & 1.3  & 0.8 & 2.4 \\ 
SE3 & 12:48:15.57 & +17:46:03.2 & 10.4 & $-17$ & $< 0.7$ & 2.7  & 0.6 & 1.1 \\ 
SW2 & 12:48:13.66 & +17:46:10.4 & 12.1 & $-20$ & $< 1.0$ & 1.7  & 1.0 & 3.4 \\ 
\enddata
\tablecomments{$\Delta r$ and $\Delta v$ are the projected distances
  (in kpc) and velocity shifts (relative to $\rm v \simeq
  6170$~km~s$^{-1}$ for the central region of PGC~043234). The
  remaining columns are the \hbeta, [O III]~$\lambda$5007, \halpha,
  and [N II]~$\lambda$6584 line
  luminosities of each region obtained after subtracting the stellar
  continuum using STARLIGHT models and correcting for Galactic extinction.}
\label{tab1}
\end{deluxetable}

\end{document}